# A frequency approach to topological identification and graphical modeling

Giacomo Innocenti *

June 22, 2018

## Abstract

This works explores and illustrates recent results developed by the author in field of dynamical network analysis [12, 21, 13, 20, 22, 11, 19, 10]. The considered approach is blind, i.e., no a priori assumptions on the interconnected systems are available. Moreover, the perspective is that of a simple "observer" who can perform no kind of test on the network in order to study the related response, that is no action or forcing input aimed to reveal particular responses of the system can be performed. In such a scenario a frequency based method of investigation is developed to obtain useful insights on the network. The information thus derived can be fruitfully exploited to build acyclic graphical models, which can be seen as extension of Bayesian Networks or Markov Chains. Moreover, it is shown that the topology of polytree linear networks can be exactly identified via the same mathematical tools. In this respect, it is worth observing that important real systems, such as all the transportation networks, fit this class.

# 1 Introduction

## 1.1 Problem background

In the last decade the analysis and the study of networks has become ubiquitous. In particular, in the study of complex systems the comprehension of the internal connections among the elementary units, which define the underlying structure of the process, turns out to play a key role to fully understand its principal mechanisms. While networks of dynamical systems have been deeply studied and analysed in systems theory (see [27, 16, 17] and references within for a sample of recent studies), a great need still exists for methods to infer their structure and their topological characteristics [23]. Indeed, in most engineering scenarios the network is given or it is the very objective of design. On the other hand, dynamical networks represent a powerful tool to model large scale systems, dividing the problem into a set of reduced complexity subsystems to be designed. In this scenario inferring a suitable internal topology is of fundamental importance. In other situations the structure of the network can not be a priori known, but its agents can be designed to identify what their neighbours are, to interact with them following a distributed strategy and, more generally, to establish any sort of consensus [26]. In this case, typical examples are given by sensor networks or cooperative controllers [25]. On the other hand, there are also interesting situations where the agents are not directly manipulable, the link structure is actually unknown and it is important to reconstruct it along with the underlying dynamics.

In this perspective, Graphical Models are a marriage between probability theory and graph theory. They provide a natural tool for dealing with two problems that occur throughout applied mathematics and engineering – uncertainty and complexity – and in particular they are playing an increasingly important role in the design and analysis of Machine Learning algorithms and Neural Networks. Fundamental to the idea of a graphical model is the notion of modularity, i.e., a complex systems turn out by combining simpler parts. Probability theory provides the glue whereby the parts are combined, ensuring that the system as a whole is consistent, and providing ways to interface models to data. Many of the classical multivariate probabilistic systems studied in fields such as Statistics, Systems Engineering, Information theory, Pattern Recognition and Statistical Mechanics are special cases of the general graphical model formalism; examples include mixture models, factor analysis, hidden Markov models, Kalman filters and Ising models. The graphical model framework provides a way to view all of these systems as instances of a common underlying formalism.

*The author is with Dipartimento di Ingegneria dell'Informazione, Università di Siena, Via Roma 56, 53100, Siena and with Dipartimento di Sistemi e Informatica, Università di Firenze, Via di S. Marta 3, 50139, Firenze, e-mail: `innocenti@dii.unisi.it`

Relevant examples about the importance of the connection structure in defining the properties of a system can be observed in different fields, such as Economics [18], Biology [9, 8], Ecology [33] and Neural Sciences [2]. We commonly find this kind of problems in Biological Neural Networks [2], Biochemical Metabolic Pathways [8], Ecology [33] and also Financial Markets (especially when trade is happening at a high frequency) [24, 32]. In all these cases very little is known about the network topology and the dynamics among the elementary agents. Hence, even inferring some relation properties between two different nodes constitutes a precious information [5].

To the best knowledge of the author, there are only few theoretical or methodological results about the reconstruction of an unknown topology related to a set of networked dynamical systems when the agents are not manipulable. Recently, this challenging problem has started drawing the attention of many researchers, especially in the physics community. Interesting and novel results are given in [31], [23] and [5], even though they are not systematic and do not provide theoretical guarantees about the correctness of the identified links and the related dynamics. In particular, in [31], [23] and [5] both theoretical and numerical procedures to detect the active links inside a dynamical networks are developed for specific classes of models. In [23] the attention is focused on sparse networks affected by noise and the aim is obtaining a final topology with a reduced number of links. No guarantee of an exact identification is provided. In [31] a more general class of networks is addressed, but the proposed method to identify their topology is based on the assumption that the input of every single node can be manipulated and that it is possible to perform many experiments to detect the adjacent links. Unfortunately, such an hypothesis is not feasible in most real-world situations and it is not practical for large networks. In[5] a generalization of Granger causality is introduced to quantify the amount of shared information among times series.

Even though the above mentioned works can be settled in the problem of reconstructing the internal connections of a dynamical network, these approaches are not yet systematic and they do not provide theoretical guarantees about the correctness of the identified links and, possibly, of the corresponding dynamics.

In this paper, the results developed in the recent years by the author in the above field are presented and summarized. A short introduction of the reference scenario and the main assumption and ideas are reported hereafter.

## 1.2 Blind frequency approach

In the following a method to infer a simplified connected graph, based on standard tools from identification and filtering theory [14, 1], is presented. The proposed approach is blind, in the sense that no a priori knowledge about the original network topology is given. Moreover, every observed process is associated with a distinct agent and only their outputs can be measured. Furthermore, no action can be performed to alter the network behaviour in order to test its response to external perturbations. Then, we assume to measure the scalar outputs $\{X_j\}_{j=1,...,N}$ of a network system composed of many interacting agents without knowing what the internal interconnections are and without any possibility to alter the functioning via a sort of testing input.

The main idea is to estimate each signal by using the others as the inputs of a dynamical system to be identified. Indeed, the corresponding models provide a description of the dependencies among the processes and of the internal dynamics of the whole system, as well. Moreover, for each agent, every process acting as input in its dynamical model can be associated to a link in the reconstructed network topology. Then, the optimal solution in terms of the minimization of the modeling error would be given by an a priori large number of dynamical models, namely one for every possible link. However, this approach is not useful because of both the a priori large number of model to be identified and the lack of proper distinctions among the links.

The proposed method, instead, aims to reconstruct a sparse link structure, that is, to select only a limited number of chosen inputs. Such a purpose is motivated by the need to provide a final model able to provide immediate information about the network structure by detecting the "most important" links. This approach is suggested by the observation that in a large network one could expect that most links could be neglected without affecting in a significant way the quality of the identification. Consequently, suboptimal strategies not only need to be taken into account, but can also be more informative to detect the internal mechanisms of interdependence among the signals. Indeed, one may obtain a better understanding of the network topology by forcing a proper selection of a reduced number of links.

To this aim, we show how a simple suboptimal strategy can be employed to provide a graphical model of the network with a reduced complexity, where the links associated to each agent could also be used to single out a selection of inputs for its dynamical model. The main goal is to obtain a simplified model of the network, which can be used to reveal some insights about its internal structure. Therefore, given a set of $N$ signals $\{X_j\}_{j=1,...,N}$, the problem of identifying, for each of them, the $m_j$ signals providing the "best" estimate is defined. This optimization problem is combinatorially intractable and it can be approached only for small values of $m_j$. In this paper, however, it is shown that solving the problem in the simplest case, where $m_j = 1$ for any $j$, provides information about the connection structure of the processes. Such a result also suggests to exploit the corresponding topology to derive for each $j$ a suboptimal set of $m_j$ processes, possibly with $m_j \geq 1$, in order to estimate via multivariate linear filtering the dynamics of the related $X_j$. We also examine some general properties of the network structure, which can be obtained in this special case, and we investigate how it can be employed as a useful tool to unveil the network structure gaining insights into its topology. Such an information not only provides a deeper understanding of the internal dynamics, but it also represents a useful tool in the realization of a suitable model of the whole system. A class of dynamical networks that can be completely and exactly disclosed via the proposed method is also present. Such a class turns out quite common and, for instance, it contains distribution networks. The application of the above approach to clustering problems will also be presented.

## 1.3 Notation

Let $\mathbb{R}$ and $\mathbb{C}$ respectively be the real and complex numbers sets. Then, for any $z \in \mathbb{C}$ such that $z = \alpha + j\beta$, with $\alpha, \beta \in \mathbb{R}$, the complex conjugate of $z$ is defined as $z^* = \alpha - j\beta$.
Let $X(t)$ and $Y(t)$ be time-discrete zero-mean wide-sense stationary processes. We denote by $E[\cdot]$ the mean operator. It follows that $E[X(t)] = E[Y(t)] = 0$. Since the two processes are wide-sense stationary it is possible to define the cross-covariance and the covariance functions, $R_{XY}(\tau) := E[X(t)Y(t+\tau)]$ and $R_X(\tau) := R_{XX}(\tau)$, with no ambiguity since there is no dependence on $t$. Given a time-discrete sequence $a(t)$, we also define

$$\mathcal{Z}[a(\cdot)](z) := \sum_{t=-\infty}^{+\infty} a(t) z^{-t}$$

as the bilateral Zeta-transform of the signal $a(t)$ for any $z \in \mathbb{C}$ such that the above sum exists. Moreover, we denote by $\Phi_{XY}(z) := \mathcal{Z}[R_{XY}(\cdot)](z)$ the cross-power spectral density between $X(t)$ and $Y(t)$ and by $\Phi_X(z) := \Phi_{XX}(z)$ the power spectral density of $X(t)$ assuming that the bilateral Zeta-transform converges in a neighborhood of the unit circle $|z| = 1$. Finally, let $\phi_X(\omega) = \Phi_X(e^{i\omega})$ and also let

$$\{\mathcal{Z}[a(\cdot)](z)\}_C := \sum_{t=0}^{+\infty} a(t) z^{-t}$$

be the causal truncation of a filter or a signal.

# 2 Problem set up

Proofs and mathematical details are omitted for the sake of simplicity. See [12, 21, 10, 11] for a further reading.

## 2.1 Mathematical tools

In this section some basic notions of graph theory, which are functional to the following developments are recalled as well as the main mathematical tools. For an extensive overview on graph theory see [7]. First we provide the standard definition of a graph.

**Definition 1.** *An* indirected *(or* undirected*) graph $G$ is a pair $(V, A)$ comprising a finite set $V$ of vertexes (or nodes) together with a set $A$ of edges (or arcs), which are unordered subsets of two distinct elements of $V$. A* directed *graph differs from the indirect type, because the edges are ordered couple of nodes, the first one being denoted as the departing node of the arc and the second the arriving vertex. The order of the couple define the direction of the edge.*

In a graph it can also be defined the notion of "path" linking two nodes as an ordered sequence of contiguous edges having them at the extremities. A path formed by coherently directed edges is said a directed path. The following definition introduces the notion of "tree".

**Definition 2.** *A graph $T = (V, A)$ is a* tree *if for any pair of distinct nodes $N_i$, $N_j \in V$ there is a unique path linking them.*

The more standard definition of a tree as an *acyclic and connected graph* is equivalent to Definition 2 as shown in [7]. However, in the development of our results we will extensively employ the property of uniqueness of a path linking two nodes. It is also important to recall the notion of connected graph, that consists in the existence of a path between every couple of its nodes.

**Lemma 3.** *Given a connected graph $G = (V, A_G)$, there exists a tree $T = (V, A_T)$ such that $A_T \subseteq A_G$. The tree $T$ is also said to be a* spanning tree *for $G$.*

The use of a weighting function on the edges of a graph allows one to define a weight for any of its "subgraphs" as the sum of the values associated to each of the corresponding arcs.

**Definition 4.** *Given a connected graph $G = (V, A)$ and a weighting function $w : A \to \Re$, we define Minimum Spanning Tree (MST) any spanning tree $T$ of $G$ such that*

$$w(T) \leq w(T') \tag{1}$$

*for any other spanning tree $T'$ of $G$.*

In this work directed graphs are also addressed and so the concepts of "rooted tree" and "polytree" are also introduced hereafter.

**Definition 5.** *A* rooted tree *is a directed graph $T = (V, A)$ such that its edges defines a tree and they satisfy the condition that only a unique node $N_r \in V$, referred to as the* root*, has exclusively departing edges. A* polytree *extends the rooted tree concept in the sense that the property of only departing edges is satisfied by a finite subset $\mathcal{R}$ of the node set.*

The presence of root nodes denotes a partial ordering among them in terms of couples of nodes linked by directed paths. For instance, if two vertexes are connected by a directed arc, the departing one is said a *parent* of the other, that in turn is denoted as a *child* of the first node. In the following the set of all the parents of the $j$-th node will be addressed as $\Pi(j)$. The notions of ancestor and descendant extend and generalize the above situation.

The previous definitions and lemmas are functional to introduce a rigorous model to address noisy linear dynamical systems interconnected to form a tree-like topology.

**Definition 6.** *Consider the triple $\mathcal{T} = (T, \mathcal{X}, \mathcal{H})$ where*

- *$T = (V, A)$ is a polytree counting $n$ nodes $\{N_j\}_{j=1,...,n}$ and $k$ roots among them;*
- *$\mathcal{X} = \{X_i\}_{i=1,...,n}$ is a set of $n$ zero-mean wide-sense stationary and ergodic discrete stochastic processes;*
- *$\mathcal{H} = \{H_{ji}(z)\}_{j,i=1,...,n}$ is a set of $n$ time-discrete SISO transfer functions $H_{ji}(z)$ represented in the domain of the $\mathcal{Z}$-transform and functionally linking the output of the $i$-th node with the input of the $j$-th one; it is also assumed that $H_{ji}(z) = 0$ if and only if there is no connections between the $i$-th and the $j$-th system.*

*For $j = 1, ..., n$, define the following stochastic processes*

$$\begin{cases} \varrho_j := X_j & \text{if } N_j \in \mathcal{R} \\ \varrho_j := X_j - \sum_{i \in \Pi(j)} H_{ji}(z) X_i & \text{if } N_j \text{ is child of } N_i \text{ for each } i \in \Pi(j). \end{cases} \quad (2)$$

*We say that $\mathcal{T}$ is a Acyclic Linear Network (ALN) if the following condition is satisfied*

$$E[\varrho_j \varrho_i] = 0 \quad \text{when } i \neq j. \quad (3)$$

It is worth underlining that a number of real world systems can be represented as ALNs despite the peculiar link structure of such a model. For instance, a special class of ALNs is represented by the transportation systems. A special subclass, in particular, consists in the power distribution network [].

## 2.2 ALN perspective

Let us assume that the observed system can be seen as a set of interconnected processes. Then, let us represent such a system as a set of $N$ agents interacting together according to an unknown structure. Moreover, assume that it is possible to obtain scalar measurements from these agents in the form of $N$ scalar time series. Furthermore, let it be possible to remove any deterministic component from these time series, in order to obtain $N$ stochastic processes $\{X_i\}_{i=1,...,N}$ which are wide sense stationary and with zero mean [30]. We intend to derive graphical model in term of a ALN representing the strongest interconnections among the processes in order to obtain a simplified representation of their unknown connectivity structure. This can be used as starting point for developing more complex topological descriptions, but the links of each agent, representing its strongest dependences, can also be exploited to design a suitable dynamical model for it.

The approach we follow to single out the network connectivity is given in terms of filtering theory [14]. Given two stochastic processes $X_i$ and $X_j$, we say that $X_j$ does not contain more information that $X_i$ if it is possible to fully determine $X_j$ from the knowledge of $X_i$ by the application of a suitable operator. This observation provides us with a tool to quantify the "relatedness" of two stochastic processes in the following way: they are "related", if it is possible to provide a "good" estimate of one (according to a specific criterion) by knowing the other, or viceversa. A special situation occurs when the operator mapping $X_i$ into $X_j$ is invertible. In this paper, we develop an approach intended to be used in situations where the operator, providing the best estimate of one process given the other, needs to be derived from the data itself (for example by using an identification or a match filter technique). In these cases, it is typical to adopt a parametric approach by choosing a fixed class of possible models. In this paper we limit ourselves to the class of linear time-invariant operators mapping stationary processes into stationary ones and we consider a least square criterion for the optimization. Indeed, in such a case the solution of the best estimate is well-known and it is given by the Wiener filter. We examine two scenarios: non-causal and causal Wiener filtering. In the non-causal scenario the considered transformations are always invertible. This property will lead to the definition of a (pseudo)-metric on the set of stationary processes. In the causal scenario, instead, this property is lost, but the general approach to determine an acyclic connectivity structure can still be applied.

According to the above formulation, we would like to develop a suitable graphical model describing the connectivity structure of the network. In particular, we intend to link each node to the most "related" ones, where the "relatedness" criterion comes from SISO Wiener filtering along with the acyclic topology requirement. We highlight that a similar and more general result can be obtained by modeling every single process $X_j$ as the output of a MISO (Multiple Input Single Output) linear system generalizing (13),

$$X_j(z) = e_j(z) + \sum_{i \in I_j} W_{ji}(z) X_i(z),  \qquad (4)$$

whose set of inputs $I(j)$ are the processes $\{X_\tau\}_{\tau \in I(j)}$ providing, according to a chosen criterion, the best trade-off between their number and the largest reduction of the cost function (5). However, such an approach relies on several combinatorial searches in order to point out the most appropriate input set $I(j)$. Therefore, we rather aim to develop a suboptimal solution, exploiting the weight matrices defined in the previous section, to derive graphs, whose topologies could be possibly used to design the linear MISO models of each process. To this purpose, let us represent the process $X_j \in \Theta$ as the $j$-th node of the set $\mathcal{V}$. Moreover, let us weight the edges connecting the $N$ elements of $\mathcal{V}$ according to the matrix $\mathcal{W}_\mathcal{V} \in \mathbb{R}^{N \times N}$. Moreover, consider a set $\mathcal{A}(\mathcal{W}) \subset \mathcal{V} \times \mathcal{V}$ of (un)directed arcs connecting nodes of $\mathcal{V}$ and denote by $\mathcal{G} = \mathcal{G}(\mathcal{V}, \mathcal{A}(\mathcal{W}))$ the corresponding (un)directed graph. Hence, if the nodes, which are closer in the graph to the $j$-th one, are chosen so that they represent the processes having the strongest relations with $X_j$, then the input set $I(j)$ of a possible MIMO model of $X_j$ can be constructed starting from its neighbours.

## 2.3 Preliminary results

First, we need to introduce a few preliminary results which will be exploited to define a mathematical tool for the quantitative characterization of the connections among the systems of the network. The main idea developed in this section is to define a distance among time series in order to establish a useful concept of "closeness". Specifically, we will assume a process $X_i$ as the input of a linear system $W_{ji}(Z)$ whose output will be used to estimate $X_j$. The two processes will be considered "close" if it is possible to find a linear transfer function $W_{ji}$ which provides a "good" estimate of $X_j$. In this respect, our approach will rely on standard least squares techniques, namely Wiener filtering. The distance function will be defined as the variance of the suitably filtered error signal in order to guarantee the properties of a metric. In particular, we distinguish two cases, which we refer to as the "causal" and the "non-causal" scenario. We are treating these two cases separately, because they will lead to two different kinds of graphs.

### 2.3.1 Non-causal scenario

First, for the sake of completeness, we provide a formulation of the Wiener filter that will be useful in the development of our method.

**Proposition 7** (Non-causal Wiener filter). *Given a frequency weighting transfer function $Q(z)$ analytic on the unit circle $|z| = 1$, a zero-mean wide-sense stationary scalar process $Y$ and a zero-mean wide-sense stationary vector process $Z$, the Wiener filter modeling $Y$ by $Z$ is the linear stable filter $\hat{W}(z)$ minimizing the* *variance of the error* $\varepsilon_Q := Q(z)[Y - W(z)Z]$, *that is*

$$\hat{W}(z) = \arg \min_{W(z)} E[\varepsilon_Q^2]. \tag{5}$$

*Its expression is given by*

$$\hat{W}(z) = \Phi_{YZ}(z)\Phi_Z^{-1}(z) \tag{6}$$

*and it does not depend upon $Q(z)$. Moreover, the minimized cost is equal to*

$$\min E\left[\varepsilon_Q^2\right] = \frac{1}{2\pi} \int_{-\pi}^{\pi} |Q(e^{j\omega})|^2 \left(\Phi_Y(\omega) - \Phi_{YZ}^*(\omega)\Phi_Z^{-1}(\omega)\Phi_{YZ}(\omega)\right) d\omega.$$

Given two scalar stochastic processes $X_i$, $X_j$, let $W_{ji}(z)$ be the Wiener filter modeling $X_j$ from $X_i$. As a consequence of Proposition 7 we have that, for any weighting transfer function $Q_j(z)$

$$E\left[[Q_j(z)(X_j - W_{ji}(z)X_i)]^2\right] = \int_{-\pi}^{\pi} |Q_j(e^{j\omega})|^2 \left(\Phi_{X_j}(\omega) - \frac{|\Phi_{X_j X_i}(\omega)|^2}{\Phi_{X_i}(\omega)}\right) d\omega. \tag{7}$$

By choosing $Q_j(z)$ equal to the inverse of the spectral factor of $\Phi_{X_j}(z)$, that is the stable and causally invertible filter $F_j(z)$, such that [29]:

$$\Phi_{X_j}(z) = F_j^{-1}(z)(F_j^{-1}(z))^*, \tag{8}$$

we obtain

$$\min E[\varepsilon_{F_j}^2] = \int_{-\pi}^{\pi} \left(1 - \frac{|\phi_{X_j X_i}(\omega)|^2}{\phi_{X_i}(\omega)\phi_{X_j}(\omega)}\right) d\omega. \tag{9}$$

This special choice of $Q_j(z)$ makes the cost depend explicitly on the coherence function of the two processes [14]:

$$C_{X_i X_j}(\omega) := \frac{|\phi_{X_j X_i}(\omega)|^2}{\phi_{X_i}(\omega)\phi_{X_j}(\omega)}, \tag{10}$$

which is non-negative and symmetric with respect to $\omega$. It is also well-known that the Cross-Spectral Density satisfies the Schwartz Inequality. Hence, the coherence function is limited between 0 and 1. The choice $Q_j(z) = F_j(z)$ can be now understood as motivated by the necessity to achieve a dimensionless cost function not depending on the power of the signals.

Let us consider, now, a set $\Theta$ of discrete zero mean and wide sense stationary stochastic processes. For the sake of simplicity, hereafter we will neglect the signal argument $t$ or $z$, when it can't be misunderstood. The cost obtained by the minimization of the error $\varepsilon_{F_j}$ using the Wiener filter as before allows us to define on $\Theta$ the function

$$d(X_i, X_j) := \left[\frac{1}{2\pi} \int_{-\pi}^{\pi} \left(1 - C_{X_i X_j}(\omega)\right) d\omega\right]^{1/2}, \quad \forall X_i, X_j \in \Theta. \tag{11}$$

**Proposition 8.** *The function $d(\cdot, \cdot)$ as defined in (11) is a pseudo-metric.*

**Remark 9.** *Observe that (11) assumes values between 0 and 1. The first case happens only when $C_{X_i X_j}(\omega) = 1$ for any $\omega$; the second situation, instead, is related to the scenario where $C_{X_i X_j}(\omega)$ is always equal to 0, i.e., when the cross-spectral density between $X_i$ and $X_j$ never assumes values different from zero.*

For a given set $\Theta$ counting $N$ processes we also introduce the symmetric weight matrix $D_\Theta \in \mathbb{R}^{N \times N}$ such that

$$D_\Theta(j, i) := d(X_j, X_i), \ \forall X_j, X_i \in \Theta. \tag{12}$$

It is worth observing that $D_\Theta(j, i)$ expresses how much good is the linear model of $X_j$ ($X_i$) consisting of the only input process $X_i$ ($X_j$). Therefore, the information recorded in the $j$-th raw (column) allows one to sort the elements of $\Theta$ according to their capacity of describing $X_j$ in term of the linear SISO (Single Input Single Output) model

$$X_j = e_j + W_{ji}(z)X_i, \quad i = 1, \ldots, N. \tag{13}$$

In particular, it holds that $D_\Theta(j,j) = 0 \; \forall j = 1, \ldots, N$.

### 2.3.2 Causal scenario

Given two stochastic processes $X_i$, $X_j$ and a transfer function $W_{ji}(z)$, let us consider again the quadratic cost $\varepsilon_Q := Q(z)[X_j - W_{ji}(z)X_i]$. Analogously to the previous case it is possible to derive a causal linear filter minimizing (5). The causal filter providing such a minimization is referred to as the causal Wiener filter [14].

**Proposition 10** (Causal Wiener filter). *The Causal Wiener filter modeling $X_j$ by $X_i$ is the causal stable linear filter $\hat{W}_{ji}^C(z)$ minimizing the filtered quantity $\varepsilon_{Q_j} := Q_j(z)[X_j - W_{ji}(z)X_i]$. It does not depend upon $Q_j(z)$ and its expression is given by*

$$\hat{W}_{ji}^C(z) = F_j^{-1}(z) \left\{ F_j(z) \frac{\Phi_{X_i X_j}(z)}{\Phi_{X_i}(z)} \right\}_C, \tag{14}$$

*where $F_j(z)$ is the inverse of the spectral factor of $\Phi_{X_j}(z)$.*

Since the weighting function $Q_j(z)$ does not affect the Wiener filter, but only the energy of the filtered error, we can choose again $Q_j(z)$ equal to $F_j(z)$, the inverse of the spectral factor of $\Phi_{X_j}(z)$. This choice simply operates a normalization of the error spectrum at every frequency. Thus, similarly to what we have done in the non-causal framework, we can define the function

$$d_C(X_j, X_i) := \left\{ E\left[ \left(F_j(z)(X_j - \hat{W}_{ij}^C(z)X_i)\right)^2 \right] \right\}^{\frac{1}{2}} \tag{15}$$

to represent the modeling error on the process $X_j$ due to the application of the causal Wiener filter to $X_i$. To highlight the properties of the present approach, it is important to notice a difference between $d_C$ and the function $d$ defined by (11). Indeed, it is straightforward to observe that $d_C$ is not symmetric and, therefore, that it can't be a metric, providing us with a less general tool to handle the processes. However, $d_C$ still represents a modeling error and in this respect it well suits the original problem of quantifying the "relatedness" between two processes.

As in the previous case, we introduce for a given set $\Theta$ counting $N$ processes the weight matrix $D_\Theta^C \in \mathbb{R}^{N \times N}$ such that

$$D_\Theta^C(j, i) := d_C(X_j, X_i), \; \forall X_j, X_i \in \Theta. \tag{16}$$

It is worth observing that $D_\Theta^C(j, i)$ still represents the capacity of $X_i$ in describing $X_j$ according to the SISO model (13), but only via causal filters.

## 3 Graphical models

### 3.1 ALN formulation

According to the above introduced scenario, let us depict each stochastic process $X_j(t)$ as the superposition of linear dynamic transformations of the other processes' outputs:

$$X_j(z) = e_j(z) + \sum_{j=1, j \neq i}^{N} W_{ji}(z)X_i(z), \tag{17}$$

where $W_{ji}(z)$ is a suitable transfer function and $e_j$ is the model error. In this framework, it can be considered interesting to find the set of $\{W_{ji}(z)\}_{i,j=1...N, i\neq j}$ which allows us to best describe the time series according to the least squares criterion

$$\min \sum_j E\left[\varepsilon_j^2(t)\right],\qquad(18)$$

where $\varepsilon_j(z) = Q_j(z)e_j(z)$ and the $Q_j(z)$ are dynamic weight functions, i.e. transfer functions. The best description provided by (17) according to (18) could be exploited in order to infer dynamic relations between the random processes. The problem with such an approach is due to the complexity of the final model, since it may be given by an a priori large number of transfer functions, namely $N(N-1)$. Hence, it is quite natural to develop a suboptimal strategy to reduce its complexity. In particular, this boils down to the problem of choosing for every $j = 1, \ldots, N$ a set $I(j) \subset \{1, \ldots, j-1, j+1, \ldots, n\}$, in order to satisfy a certain trade-off between the number of elements in each $I(j)$ and the global cost (18) associated to the modeling errors

$$e_j(z) = X_j(z) - \sum_{\tau \in I(j)} W_{j\tau}(z) X_\tau(z).\qquad(19)$$

The simplified linear model (19) represents a suboptimal strategy with respect to considering all the possible contributions present in (17). Nonetheless, if no a priori assumptions are introduced, the choice of the most suited sets $I(j)$ may turn out a complex procedure, since several combinatorial searches are needed. Therefore, in the following we develop a simplified approach inspired by graphical models, such as Bayesian Networks (BN) and Markov Random Fields (MRF). These models consist of interconnected nodes representing random variables, which are employed to provide a statistical description of the system. In particular, in both cases the inference problem associated to each node can be solved by the knowledge of a set of conditionally related random variables, which are depicted as the adjacent nodes (MRF) or as a proper extension of this set (BN). Similarly, we intend to design a procedure to build from the orginal set of processes a graph, such that its topology has a direct interpretation in terms of the sets $I(j)$.

## 3.2 Undirected graph

Undirected arcs are well suited to represent mutual relationships, where one subject affects an other and viceversa. In this perspective, between two nodes $j$ and $i$ there exists only one edge and, then, the matrix weighting the arcs results symmetric. We choose $\mathcal{W}_\mathcal{V}$ equal to $D_\Theta$ as in (12), so that the weight of each edge is a direct expression of the modeling capability between the connected nodes according to the (13) paradigm. Then, assume that, according to the above $\mathcal{W}_\mathcal{V}$, the $i$-th node is the best neighbor of the $j$-th, while the $k$-th is the best connection for the $i$-th. Observe that the weight $D_\Theta(i,k)$ is the minimum among all the values $\{D_\Theta(i,r)\}, r = 1, \ldots, i-1, i+1, \ldots, N$, while $D_\Theta(i,j)$ may be ranked in any other position. However, even though the $j$-th node may not be the most suited one to represent by itself the $i$-th, the fact, that the $i$-th is the best choice to describe with a single process the $j$-th, is meaningful. In particular, we can read this information as the fact that the $j$-th node shares with the $i$-th a certain amount of information, which no other process can model. This idea derive from a common observation in Informatics: if two different signals are able to represent the 90% of another process, but they do describe exactly the same 90%, then there is no improvement in using them both to model it. Conversely, it would be more useful to find out a signal describing, for instance, only a 20% of the objective process, if it can explain some missing information from the first chosen signal. In such a scenario, we will take both processes $X_j$ and $X_k$ as inputs for a possible MISO model (4) of $X_i$.

Observe that the above reasoning implicitly assumes that, from the modeling point of view, the amount of shared information between two processes decreases as the length of the (weighted) path linking them increases. Figuratively, this property can be regarded as though the presence of intermediate nodes in the path weakens the exchanged information. In this perspective, then, the influence of distant nodes reaches the objective process as filtered by neighbors of this latter. Due to this fact, we want to design a connected graph, since without any a priori information about the signals nature, the best practice is to assume that every process is able to affect all the others to a certain extent. Conversely, the choice of enforcing a unique path between every couple of nodes, that is of providing an acyclic graphical model, turns out reasonable in order to keep the complexity low, according to the suboptimal solution perspective.

It is worth observing that a connected graph, such that for every couple of nodes it exists a unique path linking them and such that every node is directly attached to the best neighbour according to the weight matrix, is a Minimum Spanning Tree [7]. An alternative definition is the connected acyclic graph with the least total weight. Efficient algorithms exist for its computation once the weight matrix is given.

### 3.3 Directed graph

Directed arcs are best suited to deal with binary relationships which differ according to the chosen order of the objects. Therefore, since we are also concerned with causal dependencies among the processes, we are going to develop a directed graph, whose purpose is the same as the previous one, i.e. providing a reasonable description of the connectivity structure of the entire network. As in the previous case, moreover, the resulting graphical model can be possible exploited in order to single out a suboptimal solution for the inputs of the MISO model (4) of each process.

To this aim, recalling the previous approach, it is worth observing that $D_\Theta^C$ provides distinguished information for every couple of node, according to the choice of the input. However, in order to derive a directed graph, let us choose first $\mathcal{W}_\mathcal{V}$ such that

$$\mathcal{W}_\mathcal{V}(j,i) = \min\left(D_\Theta^C(j,i), D_\Theta^C(i,j)\right) , \qquad (20)$$

so that the weight of each edge describes the best capability of one of the linked nodes in describing the other according to the (13) paradigm, also taking into account the differences due to the causality property. Moreover, to keep track of the best input in each couple, let us define the matrix $\mathcal{H}(\mathcal{W}_\mathcal{V}, D_\Theta^C)$ such that

$$\mathcal{H}(\mathcal{W}_\mathcal{V}, D_\Theta^C)(j,i) = \begin{cases} +1 & \text{if } \mathcal{W}_\mathcal{V}(j,i) = D_\Theta^C(j,i) \\ -1 & \text{if } \mathcal{W}_\mathcal{V}(j,i) = D_\Theta^C(i,j) \end{cases}. \qquad (21)$$

Similarly to the previous case, let us apply the Minimum Spanning Tree algorithm to the weight matrix $\mathcal{W}_\mathcal{V}$ as defined in (20). Then, choose for each edge $(j,i)$ the direction from the $i$-th node to the $j$-th one, if $\mathcal{H}(\mathcal{W}_\mathcal{V}, D_\Theta^C)(j,i) = +1$, and the opposite otherwise. The resulting graph is a polytree with minimum cost [15]. Analogously to the previous case, we use the arcs to represent a input/output relationships. In particular, observe that the direction of each edge defines which node would act as the input when modeling the processes via MISO linear representations. Hence, in terms of the model (4), a suboptimal though natural choice for each $I(j)$ turns out the set of the parents nodes of the $j$-th process. Observe that according to the polytree graph some processes may miss any input, due to the causal property of the considered relationships.

The above procedures, developed to design simple graphical models of a set of processes, are summarized in Table 1. It is worth highlighting that, once such a model has been derived, the analysis of the connections of its $j$-th node can be possibly exploited in order to design a reasonable input set $I(j)$ of a representation (4).

## 4 Topological identification

See [12] for a comprehensive reading about all the (omitted) mathematical details of this section.

### 4.1 Wiener filtering approach

In this section we rigorously investigate the reliability and correctness of the proposed graphical modeling procedure when it is employed to reconstruct the topology of a given network. To this aim, considering the above theory, we take into account the class of the ALN models as the unknown network whose topology has to be identified. Moreover, we propose a comparison between our approach and the results of the application of the standard MISO Wiener theory, which is briefly summarized hereafter.

| Undirected graph: |
| --- |
| `1.` represent each process as a node of the graph, obtaining the nodes set $\mathcal{V}$ |
| `2.` compute the matrix $D_\Theta$ as in (12) |
| `3.` choose the weight matrix of the graph $\mathcal{W}_\mathcal{V}$ equal to $D_\Theta$ |
| `4.` compute the Minimum Spanning Tree algorithm on the weights of $\mathcal{W}_\mathcal{V}$ |
| `5.` choose as graphical model the Minimum Spanning Tree. |
| Directed graph: |
| `1.` represent each process as a node of the graph, obtaining the nodes set $\mathcal{V}$ |
| `2.` compute the matrix $D_\Theta^C$ as in (16) |
| `3.` choose the weight matrix of the graph $\mathcal{W}_\mathcal{V}$ as in (20) |
| `4.` compute the Minimum Spanning Tree algorithm on the weights of $\mathcal{W}_\mathcal{V}$ |
| `5.` for each arc choose its direction according to (21) |
| `6.` choose as graphical model the resulting polytree. |

Table 1: Procedures to design simple graphical models.

**Definition 11** (MISO Wiener filter). *Let us consider the processes $Y$ and $\{X_i\}_{i=1,\ldots,n}$ and assume that they are wide sense stationary and with zero mean. Then, consider the MISO linear filter*

$$\sum_{i=1}^{n} \hat{W}_i(z) X_i$$

*that describes the process $Y$ by $\{X_i\}_{i=1,\ldots,n}$ and that minimizes the modelling error*

$$\min_{\substack{W_i \\ i=1,\ldots,n}} E\left[\varepsilon^2\right] = \min_{\substack{W_i \\ i=1,\ldots,n}} \frac{1}{2\pi} \int_{-\pi}^{+\pi} \phi_\varepsilon(\omega) d\omega \ , \tag{22}$$

$$\varepsilon(z) = Y(z) - \sum_{i=1}^{n} W_i(z) X_i(z) \ . \tag{23}$$

*Such a filter is referred to as the MISO Wiener filter.*

Let us now consider the class of Acyclic Linear Networks (ALN), that will be used to highlight the correctness of both our graphical method and the MISO Wiener filter in identifying a given topology. Recalling the previous definition, we can see a network in such a class as the interconnection of $m$ dynamical linear SISO systems, whose output signals $\{X_j\}_{j=1,\ldots,m}$ are linked via relationships described as:

$$\begin{cases} X_1(z) &= e_1(z) + \sum_{i:X_i \in \mathcal{P}(X_1)} G_{1i}(z) X_i(z) \ , \\ \ldots \\ X_n(z) &= e_n(z) + \sum_{i:X_i \in \mathcal{P}(X_n)} G_{ni}(z) X_i(z) \ , \end{cases} \tag{24}$$

where $\mathcal{P}(X_j)$ denotes the set of the inputs of the $j$-th system. According to the convention previously adopted, let us describe the links structure of the network by means of a graph, where the $j$-th node represents the $j$-th system and its output $X_j$ and where the edge between the $j$-th and the $i$-th nodes is present, if and only if the signal $X_j$ is an input of $X_i$, or viceversa. Therefore, recalling the ALN definition, we can cast the concepts of parent and child directly in terms of input and output of a system.

**Definition 12.** *The set $\mathcal{P}(X_j)$ of the signals, which are inputs of the $j$-th one, is referred to as the set of the parents of $X_j$. The set $\mathcal{C}(X_j)$ of the signals which have $X_j$ as a parent is referred to as the set of the children of $X_j$. A signal that has no parent is said a root, while a signal without any children is said a leaf.*

Observe that in the above network (24) the disturbances $\{e_j\}_{j=1,\ldots,n}$ are a set of jointly uncorrelated, zero-mean and wide-sense stationary processes, as defined in the ALN definition. Moreover, define the transfer function set $\mathcal{G} = \{G_{ji}(z)\}_{i,j=1,\ldots,n}$ so that $G_{ji}(z) \neq 0$ if and only if $X_i \in \mathcal{P}(X_j)$. An analogous reasoning can be performed for the path concept.

**Definition 13.** *Given two signals $X_j$ and $X_k$, if it exists a (unique) sequence $\mathcal{T}_{jk} = \{\tau_i\}_{i=1,\ldots,m}$, with $\tau_1 = j$ and $\tau_m = k$, representing $m > 1$ signals $X_{\tau_i}$, such that $H_{jk}(z) = \prod_{i=1}^{m-1} G_{\tau_i \tau_{i+1}}(z) \neq 0$, then $\mathcal{T}_{jk}$ is said a (directed) path of length $l = m-1$ between $X_j$ and $X_k$ and the related $H_{jk}(z)$ is the total transfer function associated to it.*

In the following it will be useful to define $H_{ii}(z) = 1$ for any value of $i$. Because of the polytree structure, developing each $X_j$ by iteratively substituting (24) in the expression of its components, we have that the node $X_j$ itself never appears in the right hand side of the expression. In fact, the above development leads to describe every signal via disturbances only:

$$X_j = \sum_{i=1}^{n} H_{ji}(z) e_i \ . \tag{25}$$

Without loss of generality, denote $Y = X_n$ and let us formulate the MISO Wiener filtering problem (23)-(22) regarding the modelling of $Y$ via the processes $\{X_j\}_{j=1,\ldots,n-1}$. To this aim, let us define the following quantities

$$E(z) = [e_1(z) \ \ldots \ e_n(z)]^T \ ,$$

$$W(z) = [W_1(z) \ \ldots \ W_{n-1}(z)]^T \ , \quad W_n(z) = 0 \ ,$$

$$K(z) = [H_{1n}(z) \ \ldots \ H_{nn}(z)]^T = = [K_1(z) \ \ldots \ K_n(z)]^T \ , \quad H_{nn}(z) = 1$$

$$H(z) = \begin{bmatrix} H_{11}(z) & \ldots & H_{1n-1}(z) \\ \vdots & & \vdots \\ H_{n1}(z) & \ldots & H_{nn-1}(z) \end{bmatrix} \ .$$

Therefore, in the minimization problem (22) the term (23) can be rewritten as

$$\varepsilon(z) = K^T(z) E(z) - W^T(z) H(z) E(z) = E^T(z) \left( K(z) - H(z)^T W(z) \right) \ .$$

Observe that $E^T(z) \left( K(z) - H(z)^T W(z) \right)$ is a linear combination of the signals $e_j$, which are uncorrelated. Hence, the problem boils down to

$$\min_W E\left[\varepsilon^2\right] = \min_W \int_{-\pi}^{+\pi} \sum_{j=1}^{n} \left| K_j(\omega) - \sum_{i=1}^{n-1} H_{ij}(\omega) W_i(\omega) \right|^2 \phi_{e_j}(\omega) d\omega \ , \tag{26}$$

which is formally equivalent to solve the linear least square problem

$$\min_W \left\| K(\omega) - H(\omega)^T W(\omega) \right\|^2 \ , \tag{27}$$

being $\|\cdot\|$ the euclidean norm. Intuitively, then, the signals $e_j$ act as a "basis" for the processes $X_j$ and $Y$, being the incorrelation property formally analogous to orthogonality. A so particular formulation will lead to futher developing in the following sections. This latter formulation of the problem, suitably combined with concepts borrowed from Bayesian Network theory [15], can be fruitfully exploited to investigate the results obtained via direct application of the MISO Wiener filtering to the topological identification problem.

**Definition 14.** *The Markov Blanket of the node $Y$ of a polytree network is the ensemble of all the parents and children of $Y$ and all the parents of nodes which are children of $Y$.*

**Proposition 15.** *The MISO Wiener filter between $Y$ and all the other processes $\{X_j\}_{j=1,\dots,n-1}$ may have non zero components, i.e. different form the zero transfer function, only with respect to the signals belonging to its Markov Blanket.*

Then, since the parents of $Y$ have distance equal to one from the parents of its children, the following result holds.

**Corollary 16.** *The undirected version of the topology of a linear polytree network defined as in (24) can be exactly singled out via multiple applications of the MISO Wiener filter only by using also the corresponding distance matrix (12) to purge the parents of the children of the considered node.*

### 4.2 The local coherence-based approach

In the following we investigate the results obtained by the application to ALNs of the graphical modeling procedure introduced in the previous sections. Moreover, a comparison with the Wiener filtering approach will be provided in order to highlight the advantages of our method.
To this aim, let us consider a network of interconnect linear systems of the form (24), also assuming that their link structure is described by a polytree topology. Consider the non-causal distance matrix (12) and for each possible tuple $(i,j,k)$ assume that it exists an arbitrary small interval $I$ such that

$$\phi_{X_i X_j}(\omega)\phi_{e_k}(\omega) \neq 0 \quad \forall \omega \in I . \tag{28}$$

Then, the following result holds.

**Proposition 17.** *If the $i$-th and $j$-th systems are actually linked in the real network, i.e., if $X_i$ is an input of $X_j$ or viceversa, then the indirect edge connecting the $i$-th and the $j$-th nodes belongs to the MST computed via the non-causal distance matrix (12).*

**Corollary 18** (Main result). *If condition (28) is satisfied, then the MST computed on $D_\Theta$ provides the indirect topology of the original polytree network.*

It is interesting to observe that for polytree linear network the proposed technique, based on the computation of the MST with respect to the weights matrix $D_\Theta$, provide the exact identification of the indirect version of the original topology, while the application of the Wiener filter does not.

## 5 Compressive sensing

See [22] for details.

### 5.1 Preliminary considerations

Recently, in [23] and [4] interesting equivalences between the identification of a dynamical network and a $l_0$ sparsification problem are highlighted, suggesting the difficulty of the reconstruction procedure [3].

The main idea is to cast the problem as the estimation of a "sparse Wiener filter". Recalling the problem formulation in the previous sections, given a set of $N$ stochastic processes $\mathcal{X} = \{x_1, ..., x_n\}$, we consider each $x_j$ as the output of an unknown dynamical system, the input of which is given by at most $m_j$ stochastic processes $\{x_{\alpha_{j,1}}, ..., x_{\alpha_{j,m_j}}\}$ selected from $\mathcal{X} \setminus \{x_j\}$. The choice of $\{x_{\alpha_{j,1}}, ..., x_{\alpha_{j,m_j}}\}$ is realized according to a criterion that takes into account the mean square of the modeling error. The parameters $m_j$ can be a priori defined, if we intend to impose a certain degree of sparsity on the network, or a strategy for self-tuning can be introduced penalizing the introduction of an additional link if it does not provide a significant reduction of the cost. For any possible choice of $\{x_{\alpha_{j,1}}, ..., x_{\alpha_{j,m_j}}\}$, the computation of the Wiener Filter leads to the definition of a modeling error which is a natural measure of how the time series $\{x_{\alpha_{j,1}}, ..., x_{\alpha_{j,m_j}}\}$ can "describe" the output $x_j$ in terms of predictive/smoothing capability. Once this step has been performed, each system is represented by a node of graph and, then, the arcs linking any $x_{\alpha_{j,m_k}}$ to $x_j$ are introduced for each node $x_j$. At the end of this procedure a graph, representing the network topology, has been obtained.

We will show that this way of casting the problem has strong similarities with $l_0$-minimization problems, which have been a very active topic of research in Signal Processing during the last few years. Indeed, a $l_0$-minimization problem amounts to finding the "sparsest" solution of a set of linear equations [4]. Unfortunately, with no additional assumptions on the solution, the problem is combinatorially intractable [4]. This has propelled the study of relaxed problems involving, for example, the minimization of the norm-1 which is a convex problem and it is known to provide solutions with at least a certain order of sparsity [23]. We exploit the similarity of the two problems borrowing some algorithmic tools which have been developed in the area of Signal Processing adapting them to our needs. We start introducing a pre-Hilbert space for wide-sense stochastic processes, where the inner product defines the notion of perpendicularity between two stochastic processes. This allows us to seamlessly adopt well-known greedy algorithms for a (suboptimal) solution of the problem, such as Matching Pursuit (MP) or Orthogonal Least Squares (OLS), which are based on iterated projections.

## 5.2 Pre-Hilbert space

Hereafter a method to build up a pre-Hilbert space for stationary random processes is presented. The mathematical details are omitted for the sake of simplicity (see [22])

**Definition 19.** *Let $e(t) := (e_1(t), .., e_N(t))^T$ a $N$-dimensional time-discrete, zero-mean, wide-sense stationary random process defined for $t \in \mathbb{Z}$, such that, for any $i, j \in \{1, ..., N\}$, the power spectral density $\Phi_{e_i e_j}(z)$ exists on the unit circle $|z| = 1$ of the complex plane and is real-rational. Formally, we write*

$$\Phi_{e_i e_j}(z) = \frac{A(z)}{B(z)} \quad \text{for } i, j = 1, ..., N$$

*with $A(z), B(z)$ real coefficient polynomials such that $B(z) \neq 0$ for any $z \in \mathbb{C}, |z| = 1$.*
*We say that $e$ is a vector of rationally related random processes.*

**Definition 20.** *Define the sets*

$$^0\mathcal{F} := \{W(z) | W(z) \text{ is a real-rational scalar function} \quad \text{of } z \in \mathbb{C} \text{ defined for } |z| = 1\}$$
$$^0\mathcal{F}^{m \times n} := \{W(z) | W(z) \in \mathbb{C}^{m \times n} \text{ and any} \quad \text{of its entries is in } ^0\mathcal{F}\}.$$

**Proposition 21.** *The ensemble $(^0\mathcal{F}e, +, \cdot, \mathbb{R})$ is a vector space.*

**Definition 22.** *We define a scalar operation $< \cdot, \cdot >$ on $^0\mathcal{F}e$ in the following way*

$$< x_1, x_2 > := R_{x_1 x_2}(0).$$

**Proposition 23.** *The set $^0\mathcal{F}e$, along with the operation $< \cdot, \cdot >$ is a pre-Hilbert space (with the technical assumption that $x_1$ and $x_2$ are the same processes if $x_1 \sim x_2$).*

**Definition 24.** *For any $x \in {}^0\mathcal{F}e$ we denote the norm induced by the inner product as*

$$\|x\| := <x, x>.$$

We provide an ad-hoc version of the Wiener Filter (guaranteeing that the filter will be real rational) with an interpretation in terms of the Hilbert projection theorem. Indeed, given signals $y, x_1, ..., x_n \in {}^0\mathcal{F}e$, the Wiener Filter estimating $y$ from $x := (x_1, ..., x_n)$ can be interpreted as the operator that determines the projection of $y$ onto the subspace ${}^0\mathcal{F}x$.

**Proposition 25.** *Let $e$ be a vector of rationally related processes. Let $y$ and $x_1, ..., x_n$ be processes in the space ${}^0\mathcal{F}e$. Define $x := (x_1, ..., x_n)^T$ and consider the problem*

$$\inf_{W \in {}^0\mathcal{F}^{1 \times n}} \|y - W(z)x\|^2. \tag{29}$$

*If $\Phi_x(\omega) > 0$, for all $\omega \in [-\pi, \pi]$, the solution exists, is unique and has the form*

$$W(z) = \Phi_{yx}(z)\Phi_{xx}(z)^{-1}.$$

*Moreover, for any $W'(z) \in {}^0\mathcal{F}^{1 \times n}x$, it holds that*

$$<y - W(z)x, W'(z)x> = 0. \tag{30}$$

Let $e$ be a vector of rationally related processes. Consider a set $\mathcal{X} := \{x_1, ..., x_n\} \subset {}^0\mathcal{F}e$ of $n$ rationally related processes with zero mean and known (cross-)power spectral densities $\Phi_{x_i x_j}(z)$. For any given process $x_j$, with $j \in \{1, ..., n\}$, fix $m_j \in \mathbb{N}$ such that $0 < m_j < n-1$. We intend to identify the $m_j$ processes $x_{\alpha_{j,k}}$, $\alpha_{j,k} \neq j$, with $k = 1, ..., m_j$, which, filtered by suitable rational transfer functions $W_{j,\alpha_{j,k}}(z)$, provide the best estimate of $X_j$ in the sense of the mean squares. The mathematical formulation of the problem is the following:

$$\min_{\substack{\alpha_{j,1},...,\alpha_{j,m_j} \neq j \\ W_{j,\alpha_{j,k}}(z) \in {}^0\mathcal{F}}} \left\| x_j - \sum_{k=1}^{m_j} W_{j,\alpha_{j,k}}(z) x_{\alpha_{j,k}} \right\|^2, \tag{31}$$

where every $W_{j,\alpha_{j,k}}(z)$, with $k = 1, ..., m_j$, is a possibly non-causal transfer function. Given any set $\{\alpha_{j,k}\}_{k=1}^{m_j}$, the above problem is immediately solved by a multiple input Wiener filter. It is the determination of the parameters $\alpha_{j,k}$ that makes the problem combinatorial.

The signals $x_{\alpha_{j,k}} \in \mathcal{X}$ represent the $m_j$ signals with the "strongest affinity" with $x_j$ according to a least square criterion. Moreover, as previously observed, the result of this optimization problem lends itself to a natural interpretation in terms of Graph Theory. Recalling the approach of the previous sections, assume that the processes $x_j$, $j = 1, ..., n$, are represented respectively by the nodes $N_j$, $j = 1, ..., n$, in a graph. For $j = 1, ..., n$, draw an oriented edge from each of the nodes $N_{\alpha_{j,k}}$, $k = 1, ..., m_j$, to the node $N_j$. Following this procedure, the obtained edge set can be readily visualized and provides a qualitative description of the internal structure of the whole system in terms of the most relevant linear relations. If the goal is to model a complex system as a network of suitable agents, the parameters $m_j$'s, defining the maximum number of entering edges for the nodes $N_j$'s, can be used to adjust the degree of complexity of the graph. Indeed, when $m_j = n-1$ for any $j = 1, ..., n$, we expect to obtain a complete graph, that is all possible links will be present. Conversely, a different choice for the parameters $m_j$'s will lead to a reduction of complexity of the identified network model.

### 5.3 Links with compressive sensing

In this section we highlight the connections between the problem of reconstructing a topology and the compressive sensing problem. Such a connection is possible because of the pre-Hilbert structure constructed before. Indeed, the concept of inner product defines a notion of "projection" among stochastic processes and makes it possible to seamlessly import tools developed for the compressive sensing problem in order to tackle that of identifying a topology.

In the recent few years sparsity problems have attracted the attention of researchers in the area of Signal Processing. The reason is mainly due to the possibility of representing a signal using only few elements (words) of a redundant base (dictionary). Applications are numerous, ranging from data-compression to high-resolution interpolation, and noise filtering [6, 35].

There are many formalizations of the problem, but one of the most common is to cast it as

$$\min_{w} \|x_0 - \Psi w\|_2 \quad \text{subject to} \quad \|w\|_0 \leq m \tag{32}$$

where $n < p$; $x_0 \in \mathbb{R}^p$; $\Psi \in \mathbb{R}^{p \times n}$ is a matrix, whose columns represent a redundant base employed to approximate $x_0$, and the "zero-norm" (it is not actually a norm)

$$\|w\|_0 := |\{i \in \mathbb{N} | w_i \neq 0\}| \tag{33}$$

is defined by the number of non-zero entries of a vector $w$. It can be said that $w$ is a "simple" way to express $x_0$ as a linear combination of the columns of $\Psi$, where the concept of "simplicity" is given by a constraint on the number of non-zero entries of $w$.

For each $j = 1, ..., n$, define the following sets:

$$\mathcal{W}^{(j)} = \{W(z) \in {}^{0}\mathcal{F}^{1 \times n} | W_j(z) = 0\}, \tag{34}$$

where $W_j(z)$ denotes the $j$-th component of $W(z)$. For any $W \in \mathcal{W}^{(j)}$, define the "zero-norm" as

$$\|W\|_0 = \{\# \text{ of entries such that } \exists \, z \in \mathbb{C}, W_i(z) \neq 0\}$$

and define the random vector

$$x = (x_1, ..., x_n)^T. \tag{35}$$

The problem (31) can be formally cast as

$$\min_{W \in \mathcal{W}_j} \|x_j - Wx\|^2 \quad \text{subject to} \quad \|W\|_0 \leq m, \tag{36}$$

which is, from a formal point of view, equivalent to the standard $l_0$ problem as defined in (32).

## 5.4 Solution via suboptimal algorithms

The problem of network identification/complexity reduction we have formulated in this paper is equivalent to the problem of determining a sparse Wiener filter, as explained in the previous section, once a notion of orthogonality is introduced. This formal equivalence shows how the problem of identifying a topology can immediately inherit a set of practical tools already developed in the area of compressing sensing.

Here we present, as illustrative examples, modifications of algorithms and strategies, well-known in the Signal Processing community, which can be adopted to obtain suboptimal solutions to the problem of identifying a network.

While formally identical to (32), the problem of identifying a topology as cast in (36) still has its own characteristics. Since the "projection" procedure in (36) is given by the estimation of a Wiener filter, it is computationally more expensive than the standard projection in the space of vectors of real numbers.

For this reason greedy algorithms offer a good approach to tackle the problem since speed becomes a fundamental factor. Moreover, since the complexity of the network model is the final goal, greedy algorithms are a natural solution allowing one to specify explicitly the connection degree $m_j$ of every node $x_j$. This feature is in general not provided by other algorithms.

As an alternative approach to greedy algorithms we also describe a strategy based on iterated reweighted optimizations as described in [4].

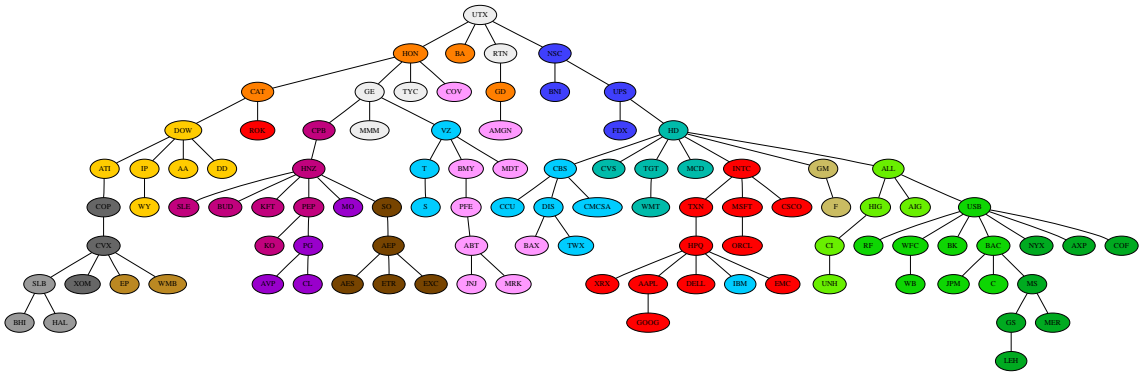

Figure 1: The tree structure obtained using a coherence-based analysis. Every node represents a stock and the color represents the business sector it belongs to. The considered sectors are `Basic Material` (yellow), `Conglomerates` (white), `Healthcare` (pink), `Transportations` (dark blue), `Technology` (red), `Capital Goods` (orange), `Utilities` (brown tints), `Consumer` (violet tints), `Financial` (green tints), `Energy` (grey tints) `Services` (light blue tints). Using the industry classification given by Google, the `Financial` sector has also been differentiated among Insurance Companies (light green), Banks (average green) and Investment Companies (dark green); `Services` have been divided in Information Technology (cyan) and Retail (aquamarina), `Consumer` in Food (plum) and Personal-care (purple); `Energy` in Oil & Gas (dark grey) and Well Equipment (light grey); Utilities in Electrical (dark brown) and Natural Gas (light brown). The complete stock list is reported in Table 2.

# 6 Application to clustering and graphical modeling

## 6.1 High-frequency stock market analysis

A collection of 100 stocks of the New York Stock Exchange has been observed for four weeks (nineteen market days), in the lapse 03/03/2008 - 03/28/2008 sampling the prices every 2 minutes. The stocks have been chosen as the first 100 stocks with highest trading volume according to the Standard & Poor Index at the first day of observation. An a priori organization of the market has been assumed in accordance with the sector and industry group classification provided by Google Finance®, that is also the source of our data. We underline that in this paper we are mainly concerned with sectors, but in some cases we have also taken into account the industry group classification to refine the results. The whole observation horizon spans the month of March. We have not considered reasonable the hypothesis that the mutual influences among the companies are stationary during this time period. However, it is worth observing that the market session data are naturally divided into subperiods, namely weeks and days and discontinuities are present. Indeed, due to the pre- and post-market sessions, there is a discontinuity between the end value of a day and the opening price of the next one. Therefore, in our analysis, we have followed the natural approach of dividing the historical series into nineteen subperiods corresponding to single days. Moreover, we have assumed the working hypothesis that the relations between the companies can be approximated with stationary ones during a single day. Hence, we have performed the multivariate analysis for each subperiod computing the coherence-based distance (11) among the stocks. Finally, we have averaged such distances over the whole observation horizon and the related results have been exploited to extract the MST, providing the corresponding market structure. We find useful to remark that the computation of the distances for small data sets is also computationally better performing. The corresponding results are illustrated in Figure 1. Every node represents a stock and the color represents the business sector or industry it belongs to. We note that the stocks are quite satisfactorily grouped according to their business sectors. We stress that the a priori classification in sectors is not a hard fact by itself and we are not trying to match it exactly. A company could well be categorized in a sector because of its business, but, at the same time, could show a behaviour similar to and explainable through the dynamics of other sectors. Actually, we would be very interested into finding results of this kind. Indeed, in those very cases, our quantitative analysis would provide the greatest contributions detecting in an objective way something which is "counter-intuitive". Thus, we just use such a priori classification as a tool to check if the final topology makes sense and if, at a general level, the coherence approach performs better. Despite this disclaim, it is worth noting that the `Financial` (green tints), `Consumer` (violet tints), `Basic Materials` (yellow), `Energy` (grey tints) and `Transportation` (dark blue) sectors are all perfectly grouped, with no exceptions. We note a subclusterization of the `Financial` sector, as well. The `Consumer` sector shows another prominent subclusterization in the `Food` (plum) and `Personal/Healthcare` (purple) industries, while the `Energy` sector presents an evident subclusterization into the `Oil & Gas` (dark grey) and `Oil Well Equipment` (light grey). The close presence of companies of a different sector and industry, `Utilities/Natural Gas` (light brown) is observed as well. The other `Utilities/Electricity` companies (dark brown) are, interestingly, a different group. We also observe a big cluster of companies classified as `Services` (light blue tints). We have differentiated them in the two industries `Retail` and `Information Technology` using two slightly different colors, respectively aquamarine and cyan. We also note the presence of three `Services` companies which are isolated from the other ones: `V` [Verizon], `T` [AT&T], and `S` [Sprint]. All of them are telephone companies. This might suggest that this industry should show at least a slightly different dynamics from the other service companies. Note also how the `Technology` sector (red) is almost perfectly grouped and how `IBM`, an IT company, even though classified as a `Services` company, is located in it. Finally, the two only automobile companies `GM` and `F` [Ford] happen to be linked together. The analysis of this four weeks of the month of March cleanly shows a taxonomic arrangement of the stocks even though the choice of a tree structure might have seemed quite reductive at first thought.

### 6.1.1 Benefits of the dynamical modeling

In this section we highlight the improvements of the dynamical approach, i.e. the coherence-based distance (11), against the static procedure, i.e. the traditional correlation-based distance used, for example, in [18]. However, the application of the correlation-based analysis to the entire data set would be meaningless, since the stocks are sampled at high frequency over a long observation horizon. Hence, to avoid the non-stationary problem, we perform the procedure in the daily subperiods, averaging the corresponding distances. Figure 2 illustrates a comparison between the "heat map" corresponding to the correlation-based distance matrix and the coherence-based one. The energy of the error associated to the dynamical modeling approach is expected to be lower than the static one. Conversely, the correlation-based distance is expected to be higher than the coherence-based metric. Such a property is highlighted in Figure 2 by the darker looking of the map (b) against map (a). Indeed, it shows that the dynamical modeling approach is able to describe a larger variety of dependencies, especially relations involving the presence of delays. The Spearman index has been evaluated from these set of data obtaining $\sigma^{sp} \simeq 0.158$.

The MST derived from those correlation-based distances is illustrated in Figure 3. At a very first look, a lesser capability of grouping the stocks according to their sectors already appears. Moreover, a further analysis reveals, for instance, that the subclusterization of the `Financial` sector highlighted in the dynamical approach is no more present in the correlation-based MST. The same happens for the subclusterizations of the `Consumer` and `Energy` sectors.

The higher consistency of the coherence-based metric can be also underlined considering the time evolution of the single distances. In Figure 4 we have considered a test node belonging to the `Technology` sector and we have reported along the time horizon both its distances with respect to other `Technology` stocks and the distances related to the `Financial` ones. As expected, the dynamical modeling approach is more consistent and robust in detecting similarities among stock time series belonging to the same sectors.

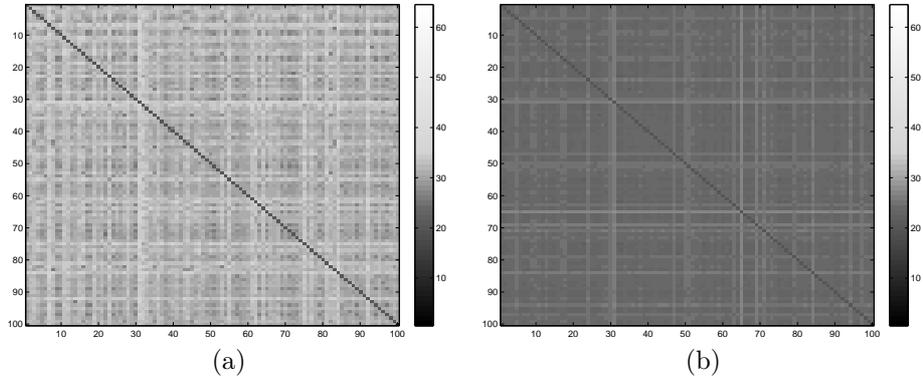

Figure 2: A graphical representation of the averaged correlation (a) and coherence (b) distance matrices. Any distance matrix has $100 \times 100$ entries representing the distances of any possible couple of the 100 stocks in our analysis. These values have been represented using a grey scale: the lighter the spot, the large the distance between the two stocks. The interpretation of both the distances in terms of a modeling error shows, as expected, a better modeling capability for the dynamic approach. This can be considered an evidence for the presence of propagative phenomena in the network, at the considered time scale of 2 minutes.

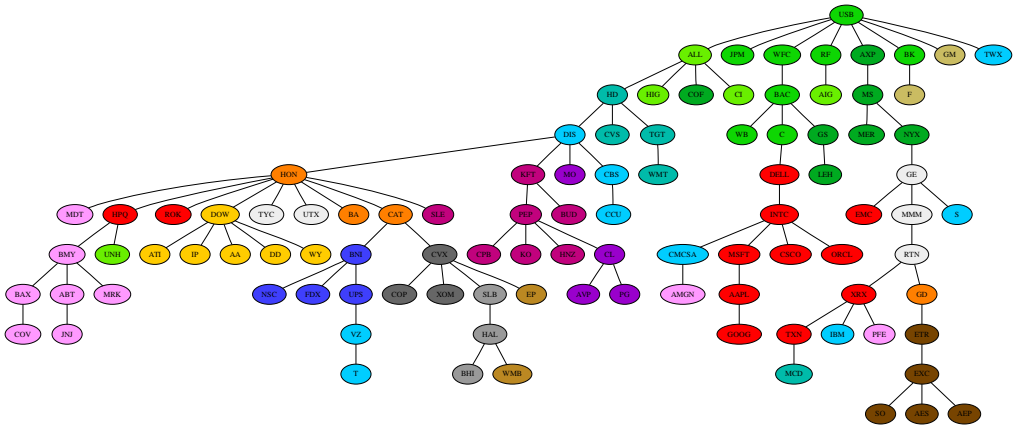

Figure 3: MST associated to the correlation-based metric. The distances have been computed in the daily subperiods and then averaged as in Figure 1. An increased number of stocks grouped with others belonging to a different sector is experienced. For instance, HPQ [Hewlett-Packard] is no more grouped with the other Technology stocks. One might also observe that GM and F are not directly linked or that the Consumer stocks (cyan) are almost spread over the whole graph. Furthermore, the subclusterization of the Financial sector, highlighted in the Figure 1, is not present.

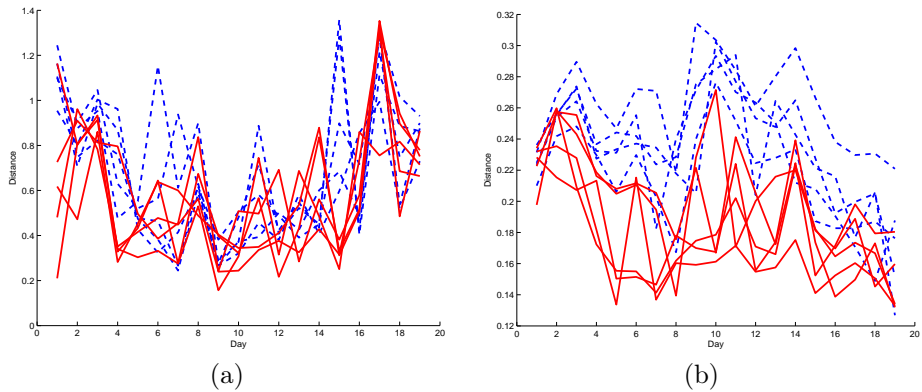

Figure 4: The distances of 5 financial stocks (blue) and 5 technological stocks (red) from a sample technological stock have been computed on a daily base. The figure represents both the correlation distances (a) and the coherence ones (b). As it can be noted, the correlation distance is more consistent and robust in detecting similarities among stocks belonging to the same sectors.

## 6.2 Graphical models of European climatic regions

In this section we provide an illustrative example to show the effectiveness of the proposed graphical modeling approach (see [12] for a further reading on this case). To this aim, we report the results obtained by analyzing a set of real data. A crucial assumption lies in the adoption of linear models for the modeling process. Indeed, the application of our technique to real data is expected to be able to provide information about the linear component of the dynamics, while the nonlinear part will necessarily be embedded as noise. In other words, the corresponding results are expected to be as worse as the nonlinear dynamics prevail in the real network. However, though a linear analysis may be not sufficient to fully understand the process dynamics, we highlight that it provides a first step to derive information about the system, especially about its internal connections. Indeed, we remind that no a priori knowledge is assumed, thus ours is a fully blind approach. In general, it is possible to incorporate a priori knowledge into the modeling procedure if, for example, the dynamics is known to have a certain structure. However, the development of similar procedures goes beyond the scope of this paper and it will be matter of future research.

In the following we address a collection of meteorological time series sampled in 50 sites located in as much European towns, listed in Table 3. Our main goal is obtaining information about the similarities among the temperatures in those towns, or, rather, deriving a suitable topological model.

The temperatures in the above locations are observed over a one year time horizon covering the whole 1988. Such time series are hourly sampled, providing a high frequency data set [1]. Notice that consequently both daily and seasonal trends may be pointed out from the observations. To this regard, any deterministic trend has to be removed from the original time series. Hence, let $\{Y_k(n)\}_{n=1,\ldots,8760}$, $k = 1, \ldots, 50$, be the hourly sampled temperatures in the considered towns. Any trend different from the daily one can be singled out just by observing the mean value over a 24 hours span, i.e.

$$S_k(n) = \frac{1}{24} \sum_{i=-12}^{11} Y_k(n+i), \quad n = 1, \ldots, 8760, \, k = 1, \ldots, 50,$$

assuming $Y_k(n) = 0$ if $n < 1$ or $n > 8760$ as a working hypothesis. Therefore, we provide the rejection of such seasonal trends just by defining the new time series

$$X_k(n) := Y_k(n) - S_k(n), \quad k = 1, \ldots, 50.$$

---

[1] International Surface Weather Observations 1982-1997, Volume 3 (Europe) , 1998 Asheville, N.C

Table 2: List of the companies considered in the analysis.

| Name | Code | Sector | Name | Code | Sector |
|---|---|---|---|---|---|
| 3M Company | MMM | Conglomerates | Goldman Sachs Group | GS | Financial |
| Abbott Laboratories | ABT | Healthcare | Google Inc. | GOOG | Technology |
| Aes Corporation | AES | Utilities | Halliburton | HAL | Energy |
| Alcoa Inc. | AA | Basic Materials | Hartford Financial Services | HIG | Financial |
| Allegheny Technologies Inc. | ATI | Basic Materials | H. J. Heinz | HNZ | Consumer/Non-Cyclical |
| Allstate Corporation | ALL | Financial | Hewlett-Packard | HPQ | Technology |
| Altria Group | MO | Consumer/Non-Cyclical | Home Depot | HD | Services |
| American Electric Power | AEP | Utilities | Honeywell International | HON | Capital Goods |
| American Express | AXP | Financial | Intel | INTC | Technology |
| American International Group | AIG | Financial | International Business Machines | IBM | Services |
| Amgen Inc. | AMGN | Healthcare | International Paper | IP | Basic Materials |
| Anheuser Busch | BUD | Consumer/Non-Cyclical | Johnson & Johnson | JNJ | Healthcare |
| Apple Inc. | AAPL | Technology | JPMorgan Chase | JPM | Financial |
| AT&T | T | Services | Kraft Foods | KFT | Consumer/Non-Cyclical |
| Avon Products | AVP | Consumer/Non-Cyclical | Lehman Brothers Holding | LEH | Financial |
| Baker Hughes Inc. | BHI | Energy | McDonald's | MCD | Services |
| Bank of America | BAC | Financial | Medtronic | MDT | Healthcare |
| Bank of New York Mellon | BK | Financial | Merck | MRK | Healthcare |
| Baxter International | BAX | Healthcare | Merril Lynch | MER | Financial |
| Boeing | BA | Capital Goods | Microsoft | MSFT | Technology |
| Bristol Myers Squibb | BMY | Healthcare | Morgan Stanley | MS | Financial |
| Burlington Northern Santa Fe | BNI | Transportation | Norfolk Souther Group | NSC | Transportation |
| Campbell Soup | CPB | Consumer/Non-Cyclical | NYSE Euronext | NYX | Financial |
| Capital One Financial | COF | Financial | Oracle | ORCL | Technology |
| Caterpillar Inc. | CAT | Capital Goods | Pespi | PEP | Consumer/Non-Cyclical |
| CBS | CBS | Services | Pfizer Inc. | PFE | Healthcare |
| Chevron | CVX | Energy | Procter & Gamble | PG | Consumer/Non-Cyclical |
| CIGNA | CI | Financial | Raytheon | RTN | Conglomerates |
| Cisco Systems | CSCO | Technology | Regions Financial | RF | Financial |
| Citigroup Inc | C | Financial | Rockwell Automation | ROK | Technology |
| Clear Channel Communications | CCU | Services | Sara Lee | SLE | Consumer/Non-Cyclical |
| Coca-Cola | KO | Consumer/Non-Cyclical | Schlumberger Limited | SLB | Energy |
| Colgate Palmolive | CL | Consumer/Non-Cyclical | Southern | SO | Utilities |
| Comcast | CMCSA | Services | Sprint Nextel | S | Services |
| Conoco Phillips | COP | Energy | Target | TGT | Services |
| Covidien | COV | Healthcare | Texas Instruments Inc. | TXN | Technology |
| CVS Caremark | CVS | Services | Time Warner | TWX | Services |
| Dell Inc | DELL | Technology | Tyco International | TYC | Conglomerates |
| Dow Chemical Company | DOW | Basic Materials | U. S. Bancorp | USB | Financial |
| E.I. du Pont de Nemours | DD | Basic Materials | United Parcel Service | UPS | Transportation |
| El Paso | EP | Utilities | United Technologies | UTX | Conglomerates |
| EMC | EMC | Technology | UnitedHealth Group Inc. | UNH | Financial |
| Entergy | ETR | Utilities | Verizon Communications | VZ | Services |
| Exelon | EXC | Utilities | Wachovia | WB | Financial |
| Exxon Mobil | XOM | Energy | Wal-Mart Stores | WMT | Services |
| FedEx | FDX | Transportation | Walt Disney | DIS | Services |
| Ford Motor | F | Consumer Cyclical | Wells Fargo | WFC | Financial |
| General Dynamics | GD | Capital Goods | Weyerhaeuser Company | WY | Basic Materials |
| General Electric | GE | Conglomerates | Williams Companies | WMB | Utilities |
| General Motors | GM | Consumer Cyclical | Xerox | XRX | Technology |

This technique to detrend data is analogous to the procedure reported in [28]. It is worth noticing that the new time series means are approximately equal to zero, since the temperature variation over a 24 hours time span is fairly distributed around the mean value, as expected by such time series. Observe that also this daily trend should be assumed as a deterministic periodic component, but, due to the nature of the data, its range is expected to vary along the year. Theoretically, we could apply a "windowed" approach to point out how the daily behavior changes, but the choice of the related time length should be derived by the seasonal trends. Therefore, for the sake of the simplicity, in this example we just neglect to remove the daily trend, considering it as part of the stochastic variation of the signal.

In the following analysis we consider a given data set and we do not require any on-line computation. Moreover, only the non-causal approach will be presented, since the causal one just differs for the numerical tools used to compute the distance matrix.

The coherence functions of the detrended time series $\{X_k\}_{k=1,\ldots,50}$ are numerically evaluated by employing the Welch algorithm [34]. Then, the corresponding distances are computed according to (11), providing a quantitative description of similarities and connections among the temperatures in the considered towns during the year 1988. Such information can be exploited to choose the best source of each time series according to the minimum modeling error approach and to derive a connected tree topology according to the procedure summarized in Table 1. The corresponding graphical model is illustrated in Figure 5.

Even though the approach is blind and it does not exploit any a priori knowledge, the modeled topology absolutely turns out reasonable. Towns belonging to the same climatic region are correctly linked together, also forming consistent clusters in term of the links associated only to the minima of the rows of the distance matrix. Moreover, some connections representing the propagative effects of the weather are provided, as well. Hence, the corresponding topology appears consistent with the problem of deriving a reduced optimal model of the whole system. For instance, observe that, according to our approach, in order to derive a suitable reduce linear model to describe the temperatures in Edinburgh, we should consider a MISO linear filter with the temperatures of Aberdeen, Carlisle and Glasgow as inputs.

| Aberdeen | ABR | Bordeaux | BRX | Berlin | BRL | Barcelona | BRC |
|---|---|---|---|---|---|---|---|
| Belfast | BLF | Brest | BRE | Bremen | BRE | Granada | GRN |
| Birmingham | BRM | Chartres | CHR | Dresden | DRS | Madrid | MDR |
| Bournemouth | BMT | Dijon | DJN | Dusseldorf | DSS | Malaga | MLG |
| Bristol | BRS | Lyon | LYN | Frankfurt Am Main | FRN | Salamanca | SLM |
| Cardiff | CRD | Marseille | MAR | Hamburg | HMB | Santiago De Compostela | SDC |
| Carlisle | CRL | Montpellier | MNT | Hannover | HAN | Sevilla | SVL |
| Edinburgh | EDN | Nancy | NCY | Heidelberg | HDL | Valencia | VAL |
| Exeter | EXE | Nantes | NAN | Kiel | KIE | Zaragoza | ZAG |
| Glasgow | GSW | Nice | NIC | Leipzig | LPZ | | |
| London | LND | Orleans | ORL | Mannheim | MNH | | |
| Plymouth | PLY | Paris | PRS | Munich | MUN | | |
| Manchester | MAN | Strasbourg | STR | Stuttgart | STG | | |
| Nottingham | NOT | Toulouse | TOU | | | | |

Table 3: The 50 European towns considered in Section 6.2 and their labels.

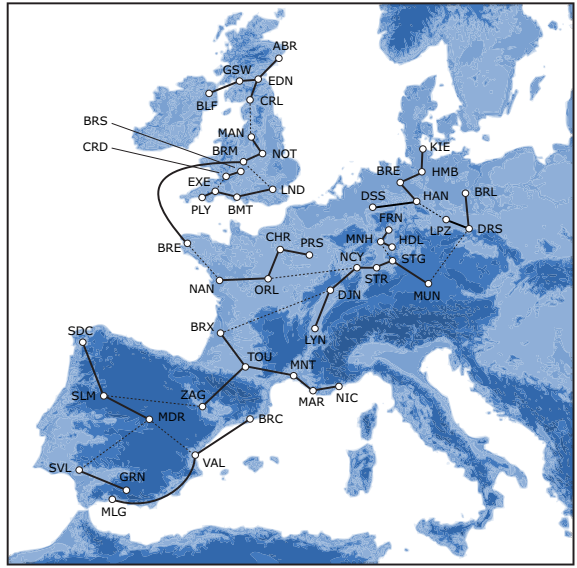

Figure 5: Solid plus dashed lines: the tree topology obtained by the application of the undirected procedure of Table 1 for the towns reported in Table 3. Solid lines: connections corresponding to the minima of each row of the undirected distance matrix.